\documentstyle[12pt]{article}
      
      \def\be{\begin{equation}}
      \def\ee{\end{equation}}
      \def\bea{\begin{eqnarray}}
      \def\eea{\end{eqnarray}}

      \renewcommand{\theequation}{\arabic{section}.\arabic{equation}}

      \def\case#1/#2{\textstyle\frac{#1}{#2}}
      \def\k0{\kappa_{0}}

\begin{document}
      \begin{titlepage}


      \begin{center}
      \Large
      {\bf Singular instantons in higher derivative theories}

      \vspace{0.4in}
      \large{B. C. Paul$^{1*}$, S. Mukherjee$^{1**}$ and R. Tavakol$^{2***}$}

      \normalsize
      \vspace{.4in}

      $^1$ {\em Physics Department, \\
      P. O. North Bengal University, \\
      Siliguri, Dist. : Darjeeling, PIN : 734 430, INDIA}

      \vspace{0.2in}

      $^2$ {\em Astronomy Unit, \\
      School of Mathematical
      Sciences,  \\
      Queen Mary, University of London, \\
      Mile End Road,
      LONDON, E1 4NS, England}

      \end{center}

      \vspace{0.3in}

      \baselineskip=24pt
      \begin{abstract}
      \noindent
      We study the Hawking-Turok (HT) instanton solutions which have been
      employed to describe
      the creation of
      an open inflationary universe, in the context of higher derivative
      theories.
      We consider the effects of adding quadratic and cubic terms of
      the forms $\alpha R^{2}$ and $\beta R^{3}$ to the
      gravitational action.
      Using a conformal transformation to convert the higher derivative theories
      into theories of self interacting scalar fields minimally coupled to
      Einstein gravity, we argue that the cubic term
      represents a generic perturbation of the polynomial type
      to the action and obtain the
      conditions on the parameters of these theories for the existence of
 singular
      and non-singular
      instanton solutions. We find that, relative to the quadratic case,
      there are significant changes in the nature of the constraints
      on the parameters for the existence of these instantons,
      once cubic (and higher order
      perturbations) are added to the action.
      
      \end{abstract}

      PACS NUMBERS: 04.20.Jb, 98.80.Hw

      \vspace{.3in}

      $^*$Electronic address: bcpaul@iucaa.ernet.in

      $^{**}$Electronic address: sailom47@hotmail.com

      $^{***}$Electronic address: r.tavakol@qmw.ac.uk
      \end{titlepage}


      \pagebreak
      \section{Introduction}

      \setcounter{equation}{0}

      \def\theequation{\thesection.\arabic{equation}}

      It is well known that the creation of  an open homogeneous universe
      requires either very special potentials \cite{Coleman-DeLuccia} or
      fine tuned initial
      conditions \cite{Linde}.
      Thus, for example, the early mechanism suggested by
      Coleman and De Luccia \cite{Coleman-DeLuccia},
      requires a special class of inflaton effective
      potentials which have a metastable minimum followed
      by a small slope region which
      permits a further slow-roll inflation. The inflaton field is supposed to be
      initially trapped in the false vacuum leading to a period of inflation,
      which gives an almost de Sitter space with small quantum fluctuations.
      The field eventually undergoes a quantum tunnelling, nucleating
      bubbles within which it slowly rolls down to the true
      vacuum.  The interior of such a bubble is actually an open universe.
      To keep the quantum fluctuations small, a very flat
      potential is required while one also requires a metastable minimum.
      Thus this scenario can only be realised at the cost of having
      very special potentials and
      fine tunings.

      An interesting alternative mechanism was suggested recently by
      Hawking and Turok (HT) \cite{Hawking-Turok,Turok-Hawking},
      which aims to remove
      some of these shortcomings. The HT mechanism makes use of the
      no--boundary proposal of Hartle and Hawking \cite{Hartle-Hawking}
      and provides a novel technique
      for creating an open inflationary universe  described by a singular
      instanton obtained in a minisuperspace model
      with an inflaton field.
      The HT instanton provides a method of calculating the
      probability of the creation of a homogeneous open universe
      while removing the requirement of a false vacuum
      and at the same time giving rise to a universe with
      small quantum fluctuations.
      Despite its novelty, several
      aspects of this mechanism have been criticised.
      For example Vilenkin \cite{Vilenkin}
      has criticised the use of a
      singular instanton and has produced a counter-example,
      whereby
      a model with a minimally
      coupled massless scalar field  which permits a singular instanton has been
  shown
      to lead to catastrophic results.
      The proposal and this controversy have led to considerable
      work in this field. Gratton and Turok \cite{Gratton-Turok} have argued
      that singular instantons should not be ignored, as suggested by Vilenkin,
      as the field configurations contributing to the path integral are
      in any case non-differentiable. Wu \cite{Wu} has pointed out that the
  singular
      instantons should be treated as constrained solutions which are not the
      stationary points of the gravitational action. Turok \cite{Turok}
      has further suggested
      that a careful treatment of these instantons may remove the instability
      pointed out by Vilenkin. Unruh \cite{Unruh}
      has shown that the instanton creates a
      finite bounded universe of which the homogeneous hyperbolic geometry is a
      part. Linde \cite{Linde} has
      suggested the introduction of a change of sign in the Euclidean action for
      an instanton in quantum cosmology,  giving the probability $ {\it P} \sim
      e^{\it S} $, different from what one obtains with the proposal of Hartle
 and
      Hawking. Bousso and Hawking
      \cite{Bousso-Hawking} have shown that with Linde's prescription
      the creation of a universe with large numbers of black holes is favoured
      resulting in a universe without a radiation dominated era. Further work on
      open inflation has also been done in the context of scalar-tensor gravity
      theories \cite{Lee-etal} and in
      higher dimensions. Garriga \cite{Garriga} has
      given a method for obtaining HT instantons in
      four dimensions by dimensional reduction of higher dimensional non-singular
      instantons. By including a cosmological constant in the theory,
      he obtains a non-singular instanton, although eleven dimensional
      supergravity theory does not have a cosmological constant.
      Such a theory has, however, been shown \cite{Hawking-Reall}
      to permit also a  singular instanton.

      Given the potential importance of the HT scenario,
      it is of interest to study its robustness with respect to
      additional ingredients that are expected to be
      present. Here as a step in this direction, we shall
      consider the effects of including higher derivative corrections
      to the gravitational action, on the existence of
      both non-singular de Sitter and singular HT-type solutions.
      Quadratic and higher-order terms in the Riemann curvature tensor
      and its traces appear in the low-energy limit of superstrings
      \cite{Candelas-etal}, as well as when the usual perturbation
      expansion is applied to General Relativity
      \cite{barchr83,anttom86}.
      For example, it
      is known that with suitable counter
      terms viz $C^{\mu \nu \rho \delta} C_{\mu \nu \rho \delta}, R^{2}, \Lambda$
      added to the Einstein action one may obtain a perturbation theory which is
      well behaved, formally renormalizable and asymptotically
      free.
      Some singular as well as
      non-singular instanton solutions have recently been presented
      in the case of quadratic Lagrangians \cite{Paul-Mukherjee}.
      The renormalisation of higher loop contributions introduces terms
      into the effective action that are higher than quadratic
      order. Consequently it is important to also study the effects of
      these additional terms. In this paper we shall study the effects of
      including higher order terms, by looking at the effects
      of including $R^3$-contributions to the action.
      By employing the conformal equivalence of higher-order
      gravity theories with Einstein gravity coupled to matter fields, we
      argue that this term is prototypical of the higher order
      terms than quadratic, and in this sense
      represents a more general perturbation to the
      Einstein--Hilbert action than the $R^2$-correction, at least within
      the context of four-dimensional FLRW space-times.
      This allows us to find the range of parameters in each
      theory for which the singular HT as well as non-singular instantons
      exist in these more general settings.

      The outline of the paper is as follows. In section 2, we obtain the
      conditions for the existence of
      de Sitter instanton solutions in higher derivative theories,  in particular
      in  quadratic and cubic theories.
      Section 3 contains an analogous study in the
      presence of a scalar field, using the equivalence picture,
      obtained by a conformal transformation of the
      higher order gravitational Lagrangians. Included are  also the conditions
      for the existence of
      both singular and the non-singular instantons permitted in the
      two theories, as well a
      comparative study of the two.
      Finally section 4 contains our conclusions.
      \section{Instantons in higher derivative theories}
      We consider a theory described by the following Euclidean action
      \begin{equation}
      \label{action0}
      I_{E} =  - \frac{1}{16\pi} \int f(R) \; \sqrt{g} \; d^{4} x \;
      - \frac{1}{8\pi} \int_{\partial M} K f'(R) \sqrt{h} \; d^{3} x
      \end{equation}
      where $ g $ is the determinant of the 4-dimensional metric,
      $f$ is a differentiable function of the scalar curvature $R$,
      $h_{ij}$ is  the metric induced on $\partial M$, $K = h^{ij}
      K_{ij}$ is the trace of the second fundamental form,
      and prime denotes differentiation with respect to
      the argument of the function. Here we shall
      mainly be concerned with the case where
      $f(R)$ is given by the general polynomial function
      \begin{equation}
      \label{poly-action}
      f(R) = \sum_{n=0}^{N} \lambda_n R^n.
      \end{equation}
      Choosing $ N = 2 $, one
      obtains the quadratic Lagrangian which, as was mentioned above,
      is known to have a number
      interesting features.
      Such quadratic terms in the Riemann curvature tensor
      and its trace also appear in the Gauss-Bonnet combination,
      as well as in the low energy
      limit of the superstring theory.
      If one only considers the
      conformally flat space-times in 4 dimensions,
      then in the quadratic theory the Gauss-Bonnet
      combination essentially reduces to a $R^{2}$ term.
      We shall, however, also consider the general
      cases, the prototype of which is the cubic case.

      As in the work of Hawking and Turok \cite{Hawking-Turok}, we are interested
  in
      $O(4)$ symmetric Euclidean instanton solutions
      described by the metric
      \begin{equation}
      \label{O4}
      ds^{2} = d \sigma^{2} + b^{2} (\sigma) \left( d\psi^{2} + sin^{2} \; \psi
 \;
                         d\Omega^{2}_{2} \right)
      \end{equation}
      where $d\Omega^{2}_{2}$ is the metric of 2-sphere.
      The scalar curvature for this metric is
      given by
      \begin{equation}
      \label{curv-scalar}
      R = - 6 \left( \frac{\ddot{b}}{b} +    \frac{{\dot{b}^{2}}}{b^{2}} -
      \frac{1}{b^{2}} \right),
      \end{equation}
      where  a dot  denotes a  derivative with respect to $\sigma$.
      Considering $b$ and $ R $ as independent variables we
      rewrite the action (\ref{action0}), including the constraint given by
  (\ref{curv-scalar}),
      through a Lagrange multiplier $\beta$, in the form
      \[
      I_{E} = - \frac{\pi}{4} \int d \sigma
      \left[f(R) b^{3} -  \beta \left( R + 6
      \frac{\ddot{b}}{b} +
      6 \frac{\dot{b}^{2}}{b^{2}} - \frac{6}{b^{2}} \right) \right]
      \]
      \begin{equation}
      \label{action}
      - \frac{1}{8\pi} \int_{\partial M} d^{3} x \sqrt{h} \; K \; f'(R) .
      \end{equation}
      The variation of the action (\ref{action}) with respect to R gives
      \begin{equation}
      \label{beta}
      \beta = b^{3} f'(R).
      \end{equation}
      Substituting from (\ref{beta})
      in Eq.(\ref{action}) we obtain
      \[
      I_{E} = - \frac{\pi}{4} \int_{\sigma = 0}^{\sigma_{\partial M}}
      \left[ ( f(R)
      - R f'(R)) b^{3} + 6 \left (f'(R) b \dot{b}^{2}  + f''(R) \dot{R}
      b^{2} \dot{b} + b f'(R) \right )\right] d \sigma
      \]
      \begin{equation}
      \label{action2}
      + 3 \pi \; [ \dot{b} b^{2} f'(R)]_{\sigma = 0} .
      \end{equation}
      The variation of the action (\ref{action2}) with respect
      to $b$ yields
      \begin{equation}
      \label{b-eqn}
      2 \frac{\ddot{b}}{b} + \frac{ {\dot{b}}^{2} - 1}{b^{2}}  +
      \frac{f'''(R)}{f'(R)} \dot{R}^{2} + \frac{f''(R)}{f'(R)} \ddot{R} + 2
      \frac{f''(R)}{f'(R)} \frac{\dot{b}}{b} \dot{R} - \frac{1}{2}
      \frac{f(R)}{f'(R)} + \frac{1}{2} R  = 0 .
      \end{equation}
      The de Sitter instanton solution has the form
      \begin{equation}
      b = H_{o}^{-1} \sin  H_{o} \sigma
      \end{equation}
      where $H_{o}$ is a constant which upon using (\ref{curv-scalar})
      is found to be $ R_{o} = 12 H_{o}^{2} $.
      For a de Sitter instanton solution, equation (\ref{b-eqn}) therefore
      reduces to an equation for $R_{o}$ in the form,
      \begin{equation}
      \label{con1}
      \frac{f'(R_{o})}{f(R_{o})}  = \frac{2}{R_{o}},
      \end{equation}
      which in turn shows that for a polynomial action of the form
      (\ref {poly-action}), a
      de Sitter instanton will exist
      for any real solution $x = R_{o}$ of the equation
      \begin{equation}
      \label{condition}
      \sum_{n = 0}^{N} (n-2) \lambda_{n} x^{n} = 0.
      \end{equation}
      The two cases of special interest for us are the
      quadratic and cubic Lagrangians:
      \\

      \noindent {\bf I. The quadratic case}
      \\

      \noindent Considering the
      quadratic ($ N = 2$) truncation of the Lagrangian (\ref {poly-action}),
      in the form
      \begin{equation}
      f_{2} (R) = - 2 \Lambda + R + \alpha R^{2}
      \end{equation}
      the condition (\ref{con1}) for the existence of a
      de Sitter instanton
      solution becomes
      $ R = 4
      \Lambda =
      12 H_{o}^{2}$, which implies that in this case no
      such solutions exist for
      $\Lambda =0$.    \\

      \noindent {\bf II. The cubic case}
      \\

      \noindent
      Considering the
      cubic truncation ($ N = 3$) of the action (\ref {poly-action}),
      in the form
      \begin{equation}
      f_{3} (R) =
      - 2 \Lambda + R + \alpha R^{2} + \beta R^{3},
      \end{equation}
      Eq. (\ref{con1}) implies that in this
      case a de Sitter instanton will exist for any real positive
      root of the equation
      \[
      \beta x^{3} - x  + 4 \Lambda = 0.
      \]
      Note that as opposed to the quadratic case, in this case
      de Sitter instanton solutions can also exist
      for $\Lambda = 0 $, and are given by
      $ R = 12 H_{o}^{2} = \frac{1}{\sqrt{\beta}}$.
      Note also that in both the quadratic and cubic cases,
      the existence conditions for the
      instanton solutions are independent of the parameter
      $ \alpha $, the
      reason being that the quadratic term in each action satisfies
      the equation (\ref{condition})
      identically. The corresponding action, however, does depend on it.
      \\

      \noindent {\bf III. The general polynomial case}
      \\

      The existence of de Sitter instanton solutions in the
      the case of a higher order theory based on the general
      Lagrangian (\ref{poly-action}) depends on the
      algebraic equation (\ref{condition}) having
      a real solution. This raises two important
      questions. Firstly, given the existence of
      de Sitter instanton solutions in a lower order Lagrangian theory,
      what are the effects of switching on higher order
      terms? Secondly, given that as
      $N$ increases the number of roots of
      the equation (\ref{condition}) also increase, what
      is the likelihood that such an equation has a real root and
      hence a de Sitter instanton solution.

      The former question relates to the question of stability or
      fragility of the de Sitter instanton solution with respect to addition
      of higher order terms to the Lagrangians \cite{Coley-Tavakol}.
      In this connection it is clear that as $N \to \infty$, there is
      a real solution for every odd $N$ but not necessarily for
      an even $N$. More generally it is known that
      if the series expansions are convergent (representing some
      analytic function say on $\{ |z| <r \}$) and if
      $T: \{ |z| < s \}$, where $s<r$, then for large $N$, the number of real
  zeros
      of the infinite series, representing the function in
      $T$ is the same as the number of real zeros of any
      partial sum of size greater than $N$, counting zeros by
      multiplicity \cite{Marden}.
      Regarding the latter, the results by Kac \cite{Kac}
      concerning the probability of n-th order
      polynomials to have real roots  demonstrates
      that this is in fact rare, thus demonstrating that
      such theories are likely  to have  only few
      de Sitter instanton solutions (see also \cite{Coley-Tavakol},
       \cite{Barrow-Ottewill} ).
      The above two points are important to bear in mind when considering
      such solutions in more general settings where higher order
      Lagrangian terms are present.
      \\

      Now the primary criterion
      for deciding whether an instanton solution is physically
      favoured  is to compute the Euclidean action $S_E$
      \cite{Hawking-Turok}. The wave function for the system
      to the leading approximation is then given by $e^{-S_E}$.
      This allows the calculation of the probability of creation  of a
      de Sitter universe, described by these instantons.
      The Euclidean
      action, which
      is obtained by integrating over half of the $S^{4}$ is,
      for the general action (\ref{action0}) considered here, given by
      \begin{equation}
      I_{dS}  =  - \; \frac{\pi}{6 H_{o}^{4}} \; f( H_{o}^{2}  ).
      \end{equation}
      For the quadratic and cubic cases considered here, these actions take the
  explicit forms
      \begin{equation}
      ^{2}I_{dS}  = -\left[ \frac{3 \pi}{2 \Lambda} + 12 \pi \alpha \right],
      \end{equation}
      \begin{equation}
      ^{3}I_{dS}  = -\left[ \frac{2 \pi}{H_{o}^{2}} - \frac{\pi \Lambda}{2
      H_{o}^{4}}
      + 12 \pi \alpha \right]
      \end{equation}
      respectively, where $H_{o} $ is given by  $12 H_{o}^{2} = R_{o}$ and
 $R_{o}$
      is a real positive root  of the equation (\ref{condition}).
      Note that an increase in the values of  $\alpha $ and $\beta$ enhances
      the probability of the creation of the de Sitter instanton. Although the
      solution (2.9) does not depend on $\alpha $, the action cannot remain
  negative
      unless $\alpha  > - \frac{1}{8 \Lambda} $. Thus we obtain a lower bound on
      $\alpha$ for the quadratic case.  In the cubic case, if $\Lambda = 0$,
      $R = 12 H_{o}^{2} = 1/\sqrt{\beta} $, and hence a negative $\beta $ is not
  allowed in this case.
      Since $H_{o}$ is independent of $\alpha$, it is obvious that a larger
       $\alpha > 0$, will give greater probability of creation.

          To obtain an open inflationary universe, one considers the
      $O(4)$ symmetric metric (\ref{O4}) and substitutes  $ \psi = \frac{\pi}{2}
 +
      i \tau$ to obtain
      \begin{equation}
      \label{metric}
      ds^{2} = d \sigma^{2} + b^{2}(\sigma) \left( - \;
      d\tau^{2} + cosh^{2} \; \tau \; d \Omega_{2}^{2} \right)
      \end{equation}
      which is a spatially inhomogeneous de Sitter like metric. Now setting
      $\sigma = i \; t$ and $\tau = i \; \frac{\pi}{2} + \chi $, the metric
  (\ref{metric})
      becomes
      \begin{equation}
      ds^{2} = - \; dt^{2} + a^{2}(t) \left(   d \chi^{2} + \sinh^{2} \chi \;
      d \Omega_{2}^{2} \right)
      \end{equation}
      with $ a(t) = - \; i \; b( i \; t )$,
      which is a singular expanding open model.
      \\

      Our aim here is to find out the precise conditions
      under which the HT
      singular instantons \cite{Hawking-Turok,Turok-Hawking}
      are permitted
      in each case considered above.
      We shall do this in the next section
      by employing a conformal transformation
      which converts both the quadratic and cubic Lagrangian theories
      into Einstein gravity
      with a minimally coupled,
      self interacting scalar field $\phi$.

      \section{Instantons in the conformally equivalence picture}
      It is well known that the theories with higher order
      Lagrangians are conformally equivalent to Einstein gravity with a matter
      sector containing a minimally coupled,
      self interacting scalar field $\phi$,
      with an effective potential $V(\phi)$ \cite{Barrow-Cotsakis, Maeda}.
      The precise form of the self
      interaction  of the scalar field is determined by the higher derivative
      terms in the action. For
      a general polynomial Lagrangian of order $n$ of the form
      (\ref{poly-action}), the form of the  potential is extremely complicated.
      However, one can
      obtain the qualitative behavior of the potentials
      for all values of $n$ at small and large values
      of the field  $\phi $. The asymptotic behavior of
      potentials at infinity
      depends on the  combination of the highest degree of the
      Lagrangian polynomial and the dimensionality $D$ of space-time.
      In particular,
      for $D>2N$ the potential is unbounded from above, for $D=2N$ it
      flattens into a plateau and for $D<2N$ it has an
      exponentially decaying tail \cite{Barrow-Cotsakis}.
      Importantly, for $ D < 2 N$,  $V(\phi)$ remains
      qualitatively the same as $N$ increases.
      As a result, in the 4 dimensional case, the qualitative behavior of
      $V(\phi)$ does not change relative to the $R^{3}$ case even when
      higher order terms with
      $N > 3 $ are considered.
      This implies that the $n=2$ contribution is
      rather special in four dimensions, and that the $R^3$-term is in
      this sense more generic, being prototypical of higher order
      perturbations to the action.
      Thus,
      it is important to check if the results of the
      quadratic theory
      are robust with respect to the
      more general $R^3$ perturbations.
      It is also worth mentioning in this connection
      the results of \cite{vanElst-etal}
      which show that in the case of
      the cubic theories, there exists a region
      of parameter space in which neither the tunnelling nor the
      Hartle-Hawking
      boundary conditions predict a suitable inflationary
      scenario, good enough to solve the horizon
      and flatness problems, contrary to what
      is obtained in the case of quadratic theories.
      We note that another cubic term, $ R \Box R $,
      which may also be reduced to a theory
      of Einstein gravity, minimally coupled to two scalar fields, after a
      suitable conformal transformation, is not considered here.

      As was mentioned above, the theory represented by the
      Lagrangian $f(R)$ can
      be written as a scalar field theory minimally coupled to
      Einstein gravity. More precisely,
      considering a conformal transformation of the form
      \begin{equation}
      \tilde{g}_{\mu \nu} = \Omega^{2} \; g_{\mu \nu}
      \end{equation}
      where $\ln \Omega = \frac{\phi}{\sqrt{6}} = \frac{1}{2} \ln |f' (R)|$,
      allows such a general action
      to be transformed into
      \begin{equation}
      \label{conf-action}
      I = - \int \sqrt{ \tilde{g} } \left[ R( \tilde{g} ) - \frac{1}{2}
      (\tilde{\partial} \phi)^{2} - V( \phi ) \right],
      \end{equation}
      with $V(\phi) = \frac{1}{2} e^{ - 2 \sqrt{\frac{2}{3}} \; \phi} \left[
      R \frac{\partial f}{\partial R} - f(R) \right] $.
      Thus knowing
      the precise form of $f(R)$, allows the
      corresponding potential to be determined,
      in principle.
 \\

      The Einstein field equations corresponding to the action
 (\ref{conf-action})
  is derived
      using the equation  $\frac{\partial I}{\partial \tilde{g}^{ \mu \nu}} = 0$
      and the scalar field equation is obtained from $\frac{\partial I}{\partial
      \phi} = 0$. Choosing $\tilde{g}$ to be given by the metric
      \begin{equation}
      ds^{2} = d \tilde{\sigma}^{2} + a^{2} (\tilde{\sigma}) \left( d\psi^{2} +
      \sin^{2} \; \psi \; d\Omega^{2}_{2} \right),
      \end{equation}
      the corresponding field equations become
      \begin{equation}
      \label{a-eq}
      3 \left[ \frac{ a'^{2} -
      1}{a^{2}} \right] =  \left[ \frac{1}{2} {\phi'}^{2} - V( \phi ) \right] ,
      \end{equation}
      \begin{equation}
      \label{phi-eq}
      {\phi}'' + 3 \frac{a'}{a} {\phi}' =  \frac{\partial V}{\partial \phi} ,
      \end{equation}
      where  primes denote derivatives with respect to $\tilde{\sigma}$. We now
      look for non-singular de Sitter as well as singular HT instanton solutions
      of these equations.
      The two cases of interest to us here are the
      theories with quadratic
      and cubic Lagrangians.
      \newpage
      \noindent {\bf I. The quadratic case}
      \\

      In this case the Lagrangian takes the form
      $f(R) = R + \alpha R^{2} - 2 \Lambda $, with the
      corresponding scalar field
      potential given by
      \begin{equation}
      V(\phi) = \frac{1}{8 \alpha} e^{- 2 \sqrt{\frac{2}{3}} \; \phi}
      \left[  e^{ \sqrt{\frac{2}{3}} \; \phi}  - 1 \right]^{2} + \Lambda
      e^{- 2 \sqrt{\frac{2}{3}} \; \phi} .
      \end{equation}
      With a non-zero cosmological constant, the de Sitter
      instanton solution exists for $\alpha > 0$, $\Lambda > 0$ as well as for
      $\alpha < 0$, $ 0 < \Lambda < \frac{1}{8 |\alpha| } $, and
      is given by
      \[
      \phi = \phi_{o} = \sqrt{\frac{3}{2}} \ln (1 + 8 \alpha \Lambda ) ,
      \]
      \begin{equation}
      a(\tilde{\sigma}) = \sqrt{ \frac{3 (1 + 8 \alpha \Lambda)}{ \Lambda} } \sin
      \sqrt{\frac{ \Lambda}{3 (1 + 8 \alpha \Lambda)}}  \tilde{\sigma} .
      \end{equation}
      The corresponding Euclidean action evaluated in this case becomes
      \begin{equation}
      I_{E} = - \left( \frac{3 \pi}{2 \Lambda} + 12 \pi \alpha \right)
      \end{equation}
      which is the same as in the original quadratic theory.

      Considering the conditions for the singular HT instanton solutions,
      we note that such solutions
      cannot
      be obtained for all values of the parameters $\alpha $ and $ \Lambda $ in
  the
      quadratic theory.
      The existence conditions can be obtained
      by recalling the general features of the potential
      $V(\phi)$ in this
      case (with $\alpha' = - \alpha $), namely:

 \vskip 0.2in

      $\bullet$
      $V(\phi)$ has two zeros, $\phi_{+}$ and $\phi_{-}$ , given by
      \begin{equation}
      \phi_{\pm} =  \sqrt{\frac{3 }{2 }} \ln ( 1 \pm \sqrt{8 \alpha' \Lambda}).
      \end{equation}

      $\bullet$  $V(\phi)$ has a maximum at
      $\phi = \phi_{m} = \sqrt{\frac{3 }{2 }}
      \ln ( 1 - 8 \alpha' \Lambda)$, with $V(\phi_{m}) = \frac{\Lambda}{1 -
      8 \alpha' \Lambda}$.

      $\bullet$
      $V(\phi) \rightarrow - \frac{1}{8 {\alpha}'} $ as
      $\phi \rightarrow \infty$.

 \vskip 0.2in

      The choice of initial conditions play a dominant role in the evolution of
      these solutions. Let us assume that at $\tilde{\sigma} = 0$, $\phi(0) =
      \phi_{+} $ and $V(\phi_{+}) =  0$. It then follows that ${\phi}'(0) = 0$
      and $ \frac{dV}{d \phi}|_{\phi_{+}} =
      - \frac{\sqrt{\Lambda}}{ \sqrt{3 {\alpha}'}
      ( 1 + \sqrt{8 {\alpha}' \Lambda})}$. Since the point $\tilde{\sigma} = 0$
      is a non-singular point, the manifold looks locally like $R^{4}$ in
      spherical
      polar coordinates and we may assume $a(\tilde{\sigma}) = v_{o}
      \tilde{\sigma} + 0( \tilde{\sigma}^{2})$
      where $v_{o} $ is the initial velocity, i.e., $ a'(o) = v_{o}$.
      The potential has a negative gradient at $ \tilde{\sigma} = 0$. The initial
      conditions along with the field Eqs. (\ref{a-eq})
      and (\ref{phi-eq}) determine the evolutions of
      $b$ and $\phi$.

      To find conditions for the existence of
      singular HT  in this case,
      we assume that close to the singularity $\tilde{\sigma}_{f}$,
      ( $ \tilde{\sigma}_{f} -  \tilde{\sigma} < 1$ )
      behave as
      \begin{equation}
      \label{phi-ans}
      \phi = q \; \ln ( \tilde{\sigma}_{f}- \tilde{\sigma} )
      \end{equation}
      \begin{equation}
      \label{a-ans}
      a \sim  ( \tilde{\sigma}_{f}- \tilde{\sigma} )^{n}
      \end{equation}
      Eq. (\ref{a-eq}) then determines $q = \sqrt{\frac{3}{2}}$ for $n < 1$
      and
      Eq.(\ref{phi-eq}) gives
      \begin{equation}
      \label{non-sing}
      n = \frac{3}{4} = \frac{1}{3}
      \left[ 1 + \frac{1 - 8 {\alpha}' \Lambda }{6 {\alpha}'}
      \right].
      \end{equation}
      This in turn determines $\Lambda$ in terms of ${\alpha}'$
      \[
      \Lambda = f( {\alpha}' )  = \frac{ 2 - 15 {\alpha}'}{16 {\alpha}'}
      \]
      and hence results in the bound $ {\alpha}' < \frac{2}{15}$
      on the coupling constant of the quadratic term. Thus
      the conditions for the existence of a singular HT instanton
      (a singularity at which $a$ vanishes as
      $\tilde{\sigma} \to \tilde{\sigma}_{f}$ and $\phi$ diverges
 logarithmically)
      are
      in this case
      given by: $\alpha < 0,
      \Lambda = f(\alpha)$ and $ 8 |\alpha| \Lambda < 1$.
      There are therefore the following possibilities in this
      case:
 \vskip 0.2in
      ${\bullet} $ The scalar field may move uphill and become stabilised at
      $\phi_{m}$, giving the

      $~~$non-singular instanton solution.

      ${\bullet}$ The universe may end up in a singularity
      at $\tilde{\sigma} =
      \tilde{\sigma}_{f}$, giving
      the singular HT

      $~~$ instanton.
          \\

      The important point here is that the existence of
      singular HT instantons in this case requires
      $\alpha <0$, which
      makes them rather special.
      We also note that in this
      model, $V(\phi)$ cannot be neglected even close to the singularity,
      in contradiction to the case considered by \cite{Hawking-Turok}.
      \\

      \noindent {\bf II. The cubic case}
      \\

      In this case the Lagrangian takes the form
      $f(R) = R + \alpha R^{2} + \beta R^{3} - 2 \Lambda $,
      with the corresponding scalar field
      potential given by
      \[
      V(\phi) =   \frac{\alpha^{3}}{27 \beta^{2}}
                      e^{- 2 \sqrt{\frac{2}{3}} \; \phi }
      \left[ -1 + \frac{9 \beta}{2 \alpha^{2}}
      \left( 1  - e^{\sqrt{\frac{2}{3}} \; \phi } \right)
      + \left( 1 - \frac{3 \beta}{\alpha^{2}}
      \left(1 - e^{\sqrt{\frac{2}{3}}
      \; \phi } \right) \right)^{\frac{3}{2}} \right]
      \]
      \begin{equation}
      \label{pot-cubic}
      + \Lambda e^{- 2 \sqrt{\frac{2}{3}} \; \phi} .
      \end{equation}

      With a non-zero cosmological constant,
      there exist de Sitter instanton solutions in this case of
      the form
      \[
      \phi = \phi_{o},
      \]
      \begin{equation}
      a(\tilde{\sigma}) =  \frac{1}{H_{o}} \sin H_{o} \tilde{\sigma}
      \end{equation}
      where $\phi_{o}$ is the solution of $dV(\phi)/d\phi =0$ for the potential
      (\ref{pot-cubic}) and $H_{o}$ is given by $3 H_{o}^{2} = V(\phi_{o})$.
      In the case of $\Lambda =0$, the Euclidean action can be evaluated to be
      \begin{equation}
      I_{E} = - 12 \pi (\alpha + 2 \sqrt{\beta}),
      \end{equation}
      which is real and negative for $\alpha >0, \beta>0$.
 \\

  We now look for consitions for the existence of
  singular HT instanton in the cubic theory.
  We recall that the scalar field potential in this case has a maximum for
  $\alpha > 0$ and $\beta \geq 0$, but
  becomes flat at large $\phi$ for $\beta \rightarrow 0$,
  as in the quadratic theory.
  Fig.\ 1 shows the potential $V(\phi)$ (\ref{pot-cubic})
  for a number of values of $\beta$
  with $\alpha = 1$ and $\Lambda = 0$.
 Given that in this case
 $\alpha < 0$ makes $V(\phi) < 0$,
 we shall restrict ourselves to
 $ \alpha > 0 $.

  \input{epsf}
  \begin{figure}
  \epsffile{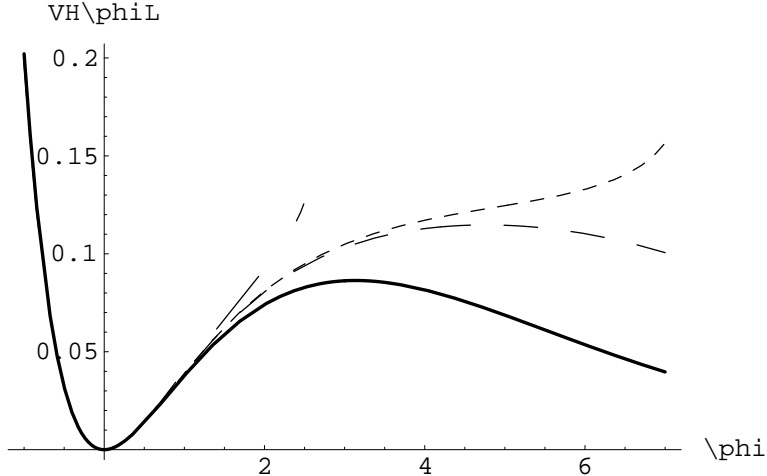}
  \caption{Plots of the potential $V(\phi)$
  (\ref{pot-cubic}), with $\Lambda = 0$ , $\alpha = 1$
  and various values of $\beta$.
  The dark, dashed,  broken  and large-dashed lines
  correspond to $\beta = 0.05,  0.002, - 0.001$ and $- 0.05$ respectively.}
  \end{figure}
  Now in the case of the cubic theory with $\Lambda = 0$, one obtains
  HT instantons corresponding to
  the solution of the equations (3.28) and (3.29) with
  \begin{equation}
      q= \sqrt{\frac{3}{2}} \; \; and \; \;
      n =  \frac{3}{4}.
      \end{equation}
      On the other hand, when $\Lambda \neq 0$, HT instantons are allowed if
      $\Lambda $ satisfies
  \begin{equation}
  \Lambda = - \frac{15}{16} - \frac{\alpha^{3}}{ 27 \beta^{2}} \left[ - 1
  + \frac{9 \beta}{2 \alpha^{2}} + \left(
  1 - \frac{3 \beta}{\alpha^{2}}
  \right)^{3/2} \right].
  \end{equation}
  Thus $\Lambda$ in this case is determined in terms of the dimensional constants
  $\alpha$ and $\beta$
  satisfying the constraint $\beta < \frac{\alpha^{2}}{3}$.
  Note that  $\alpha$ can assume some small negative values if  $\Lambda \neq 0$.
  The conditions for the existence of HT instantons in this case
  are:

 \vskip 0.2in

  $\bullet$  $V(\phi)$ has a maximum.

  $\bullet$ $\alpha^{2} > 3 \beta$.

  $\bullet $ $\Lambda$ satisfies the constraint given by (3.35)
  and could be positive, zero or negative,

 $~~$ depending upon the values of $\alpha$ and $\beta$.

      \section{Discussion}

      We have studied the existence of singular and non-singular
      instanton solutions in theories of gravity
      with quadratic and cubic Lagrangians.
      We find that in the quadratic case,
      the existence of both types of instantons
      requires a non--zero cosmological constant; whereas
      in the cubic case
      instanton solutions can also
      exist with a vanishing cosmological constant.

      To study the existence of HT instantons and clarify the special
      features of the two types of solutions, expressed the higher order
      Lagrangian theories as Einstein gravity minimally coupled
      to a scalar field, using a
      conformal transformation of the metric.
      In the case of the quadratic theory,
      the non-singular instanton corresponds to the scalar
      field sitting at the maximum
      of the potential. This gives an indication
      that the Lorentzian continuation of this
      solution ( with $\alpha < 0$ ) is
      stable, as is the case with the de Sitter
      solution in the quadratic
      theory for $\alpha < 0$ \cite{Coley-Tavakol}.
      The fact that HT instantons cannot be obtained
      with $\alpha > 0$ shows that they are rather special in the
      quadratic theory.
      There is also no HT instantons for $\Lambda < 0$ in this case.
      In the case of the
      cubic theory, HT instantons can exist for both $\alpha < 0$ and
      $\alpha > 0$ subject to a constraint which
      relates the cosmological constant to the
      dimensional constants $\alpha$ and $\beta$ in the gravitational action.

      Finally, in view of our discussion of higher order polynomial
      Lagrangians in section 2, we would expect
      our general results regarding the behaviour of the HT
      instantons, obtained using the cubic theory, to hold
      in higher order Lagrangian settings,
      although the algebraic relations among the
      dimensional constants, $\alpha$, $\beta$  {\it etc.} and $\Lambda $
      would be different at different orders.

      \vspace{0.2in}

      {\it Acknowledgement : } BCP and SM would like to thank IUCCA
      and IUCCA Reference Centre at North Bengal University for the
      use of facilities. RT would like to thank the
      IUCCA Reference Centre at North Bengal University for warm hospitality.
     
       \vspace{0.3in}


      \end{document}